\documentclass[11pt]{article}
\usepackage{graphicx}

\def\Re{{\cal R \mskip-4mu \lower.1ex \hbox{\it e}\,}}
\def\Im{{\cal I \mskip-5mu \lower.1ex \hbox{\it m}\,}}
\def\ie{{\it i.e.}}
\def\eg{{\it e.g.}}

\def\sub#1{_{\lower.25ex\hbox{$\scriptstyle#1$}}}
\def\tev{\,{\ifmmode\mathrm {TeV}\else TeV\fi}}
\def\gev{\,{\ifmmode\mathrm {GeV}\else GeV\fi}}
\def\mev{\,{\ifmmode\mathrm {MeV}\else MeV\fi}}
\def\mpl{\ifmmode M_{pl}\else $M_{pl}$\fi}
\def\mpl{\ifmmode \overline M_{Pl}\else $\bar M_{Pl}$\fi}
\def\to{\rightarrow}

\def\subw{_{\rm w}}
\def\mh{\ifmmode m\sbl H \else $m\sbl H$\fi}
\def\mch{\ifmmode m_{H^\pm} \else $m_{H^\pm}$\fi}
\def\mt{\ifmmode m_t\else $m_t$\fi}
\def\mc{\ifmmode m_c\else $m_c$\fi}
\def\mz{\ifmmode M_Z\else $M_Z$\fi}
\def\mw{\ifmmode M_W\else $M_W$\fi}
\def\mws{\ifmmode M_W^2 \else $M_W^2$\fi}
\def\mhs{\ifmmode m_H^2 \else $m_H^2$\fi}   
\def\mzs{\ifmmode M_Z^2 \else $M_Z^2$\fi}
\def\mts{\ifmmode m_t^2 \else $m_t^2$\fi}
\def\mcs{\ifmmode m_c^2 \else $m_c^2$\fi}
\def\mchs{\ifmmode m_{H^\pm}^2 \else $m_{H^\pm}^2$\fi}
\def\ztwo{\ifmmode Z_2\else $Z_2$\fi}
\def\zone{\ifmmode Z_1\else $Z_1$\fi}
\def\mtwo{\ifmmode M_2\else $M_2$\fi}
\def\mone{\ifmmode M_1\else $M_1$\fi}
\def\tb{\ifmmode \tan\beta \else $\tan\beta$\fi}
\def\xw{\ifmmode x\subw\else $x\subw$\fi}
\def\ch{\ifmmode H^\pm \else $H^\pm$\fi}
\def\lum{\ifmmode {\cal L}\else ${\cal L}$\fi}
\def\inpb{\,{\ifmmode {\mathrm {pb}}^{-1}\else ${\mathrm {pb}}^{-1}$\fi}}
\def\infb{\,{\ifmmode {\mathrm {fb}}^{-1}\else ${\mathrm {fb}}^{-1}$\fi}}
\def\epem{\ifmmode e^+e^-\else $e^+e^-$\fi}
\def\ppb{\ifmmode \bar pp\else $\bar pp$\fi}
\def\bsg{\ifmmode B\to X_s\gamma\else $B\to X_s\gamma$\fi}
\def\bsll{\ifmmode B\to X_s\ell^+\ell^-\else $B\to X_s\ell^+\ell^-$\fi}
\def\bstt{\ifmmode B\to X_s\tau^+\tau^-\else $B\to X_s\tau^+\tau^-$\fi}
\def\lamt{\ifmmode \tilde\lambda\else $\tilde\lambda$\fi}
\def\shat{\ifmmode \hat s\else $\hat s$\fi}
\def\that{\ifmmode \hat t\else $\hat t$\fi}
\def\uhat{\ifmmode \hat u\else $\hat u$\fi}

\newskip\zatskip \zatskip=0pt plus0pt minus0pt
\def\matth{\mathsurround=0pt}
\def\lsim{\mathrel{\mathpalette\atversim<}}
\def\gsim{\mathrel{\mathpalette\atversim>}}
\def\atversim#1#2{\lower0.7ex\vbox{\baselineskip\zatskip\lineskip\zatskip
  \lineskiplimit 0pt\ialign{$\matth#1\hfil##\hfil$\crcr#2\crcr\sim\crcr}}}

\def\grtsim{\,\,\rlap{\raise 3pt\hbox{$>$}}{\lower 3pt\hbox{$\sim$}}\,\,}
\def\lsim{\,\,\rlap{\raise 3pt\hbox{$<$}}{\lower 3pt\hbox{$\sim$}}\,\,}

\renewcommand{\thefootnote}{\fnsymbol{footnote}}

\begin{document} \begin{titlepage}
\rightline{\vbox{\halign{&#\hfil\cr
&SLAC-PUB-11534\cr
}}}
\begin{center}
\thispagestyle{empty} \flushbottom { {\Large\bf TeV-Scale Black Holes  
in Warped Higher-Curvature Gravity 
\footnote{Work supported in part
by the Department of Energy, Contract DE-AC02-76SF00515}
\footnote{e-mail:
$^a$rizzo@slac.stanford.edu}}}
\medskip
\end{center}

\centerline{Thomas G. Rizzo$^{a}$}
\vspace{8pt} 
\centerline{\it Stanford Linear
Accelerator Center, 2575 Sand Hill Rd., Menlo Park, CA, 94025}

\vspace*{0.3cm}

\begin{abstract}
We examine the properties of TeV-scale extra dimensional black holes (BH) in 
Randall-Sundrum-like models with Gauss-Bonnet higher-curvature terms present 
in the action. These theories naturally lead to a mass threshold for BH production 
in TeV particle collisions which could be observable at LHC/ILC. The lifetimes of 
such BH are examined and, in particular, we focus on the predicted lifetime 
differences between the canonical and microcanonical thermodynamical descriptions of BH 
decaying to Standard Model brane fields and the possibility of long-lived relics. The 
sensitivity of these results to the particular mix of fermions and bosons present in 
the Standard Model spectrum is also briefly examined. 
\end{abstract}



\renewcommand{\thefootnote}{\arabic{footnote}} \end{titlepage} 

%
%
%

\section{Introduction}

The Randall-Sundrum(RS) model{\cite {RS}} provides a geometric solution to the
hierarchy problem through an exponential warp factor whose size is
controlled by the separation, $\pi r_c$, of two 3-branes with non-zero 
tension, situated at the  $S^1/Z_2$ orbifold fixed points, which are embedded in 5-d
Anti-deSitter space, $AdS_5$. It has been shown that this interbrane distance 
can be naturally stabilized at a value necessary to produced the experimentally 
observed ratio of the weak and Planck scales{\cite {GW}}. In the original RS model, 
Standard Model(SM) matter is confined to
one of the 3-branes while gravity is allowed to propagate in the bulk. A 
generic signature of this kind of scenario is the existence of TeV-scale
Kaluza-Klein(KK) excitations of the graviton. These states 
would then appear as a series of spin-2 resonances{\cite {DHR}} that might be 
observable in  a number of processes at both hadron and
$e^+e^-$ colliders which probe the TeV-scale. Another possible RS signature is
the copious production of TeV scale black holes, though this is not a
qualitatively unique feature of the RS model as they also appear in non-warped 
scenarios{\cite {ADD}}. Here we will consider BH production in RS employing the most 
general self-consistent bulk action. 
 
One can easily imagine that this simple RS scenario is incomplete from either
a top-down or bottom-up perspective. Apart from the placement of the SM matter 
fields, we would generally expect some other `soft' 
modifications to the details of the scenario presented above, hopefully
without disturbing the nice qualitative features of the model in the gravity sector.  
It is reasonable to expect that at least some aspects of the full UV theory 
may leak down into these collider measurements and may lead to potentially 
significant quantitative and/or qualitative modifications of simple RS expectations 
that can be experimentally probed. These modifications may make their presence known 
via higher dimensional, possibly loop-induced, operators involving invariant products 
of the curvature tensor. Of course not all possible operators are allowed if we require 
that the generic setup of the RS model remain valid or be stable with respect to 
quantum corrections as we will discuss below.

One such extension of the basic RS model is
the existence of higher curvature invariants in the action which might be 
expected on general grounds from string theory{\cite {Zwiebach,Mavromatos}}, quantum 
gravity or other possible high-scale completions. A certain string-motivated class of 
such terms with interesting properties was first generally described
by Lovelock{\cite {Lovelock}} and, hence, are termed Lovelock invariants. 

These invariants are particularly important for a number of reasons. Consider extending 
the EH action by an arbitrary set of various higher curvature/derivative terms. The resulting 
equations of motion will generally be higher than second order. If the usual 5-d metric 
and set-up of the RS model are assumed to remain valid, 
this implies that derivatives of $\delta-$functions will be generated in the equations of 
motion that cannot be canceled by matter source terms. Thus, for consistency we must demand 
that the equations of motion in standard RS be no higher than second order. This is a rather 
strong restriction on what terms are allowed when generalizing the RS action. Other constraints 
may be of comparable importance but are more wide ranging beyond the RS scenario. For 
example, if one examines the 5-d propagating fields in the general case one usually obtains 
a set of massive tensor ghosts which need to be avoided. Lovelock long ago showed that these and 
other problems have a {\it unique} solution: construct the full action as a sum of Lovelock 
invariants. 

The Lovelock invariants come in fixed order, $m$, which
we denote as ${\cal L}_m$, that describes the number of powers of the curvature tensor, 
contracted in various ways, out of which they are constructed, \ie, 
\begin{equation}
{\cal L}_m \sim \delta^{A_1B_1...A_mB_m}_{C_1D_1...C_mD_m}~R_{A_1B_1}
~^{C_1D_1}.....R_{A_mB_m}~^{C_mD_m}\,,  
\end{equation}
where $\delta^{A_1B_1...A_mB_m}_{C_1D_1...C_mD_m}$ is the totally
antisymmetric product of Kronecker deltas and $R_{AB}~^{CD}$ is the
$D$-dimensional curvature tensor. When added to the EH action in any combination the 
resulting generalized Einstein equations of motion remain second order, are tachyon 
and ghost free, lead to a unitary perturbation theory due to the absence of 
derivatives of the metric higher than first in the action and to a naturally conserved 
and symmetric stress-energy tensor for the matter source fields. Because of this the 
basic RS setup is stable when these new terms are added to the action. 
For a 
fixed number of dimensions the number of the non-zero Lovelock invariants is quite restricted, 
\eg, for $D=4$, only ${\cal L}_0=1$ and ${\cal L}_1=R$, the ordinary Ricci scalar, can be 
present in the action; all higher order invariants can be shown to vanish by the 
properties of the curvature tensor. Adjusting 
the numerical coefficients of these terms we see that the resulting action is just 
the ordinary Einstein-Hilbert(EH) action of General Relativity with a cosmological 
constant. When D=5, as in the RS model, ${\cal L}_2=R^2-4R_{MN}R^{MN}+R_{MNAB}R^{MNAB}$, 
the Gauss-Bonnet(GB) invariant, can also be present in the action with an arbitrary 
coefficient, $\alpha$, with all other ${\cal L}_{m>2}=0$. Thus the GB term is the unique 
addition to the bulk EH action in the Lovelock framework in D=5; no higher order terms 
are allowed. The bulk action in the most general self-consistent RS model is now specified. 
Furthermore, it is important  
to recall that the GB term is in fact the leading (and here only allowed)stringy correction 
to the EH action{\cite {Zwiebach,Mavromatos}}. 

In principle, one might consider also adding brane terms of the GB form. However, since the 
branes live in D=4 these potential 
GB brane terms are at most surface terms as can be shown using the 
various curvature identities; they will not contribute to the equations of motion.  
The addition of the GB term to the EH action is thus 
a rather unique possibility from ($i$) the string point of view, ($ii$) our desire to 
maintain the setup and stability of the RS model and ($iii$) the nice behavior of the 
resulting equations of motion and usual structure graviton propagators. For these reasons 
we will concentrate on its influence on RS BH properties below.  

Some of the modifications of the RS model due the presence of GB terms have
been discussed by other authors (\eg, 
Refs. {\cite {KKL,Nojiri,Meissner,Cho,Cho2,Neupane,Charmousis,Brax,Gregory:2003px}). 
{\cite {KKL}} were the first to notice that the addition of the GB term to the EH action 
did not alter the basic structure of the RS model. In fact they were the first to argue 
that it is the only allowed addition of its   
kind; any other invariant in the D=5 action would lead to a destabilization of the model.  

Recently, we have begun a phenomenological examination of the effect of the presence 
of GB invariants in the action on the predictions of the RS model{\cite {Rizzo:2004rq}}. 
In that 
work we concentrated the modifications of graviton KK properties due to the GB 
term and how the value of the parameter $\alpha$ can be extracted from this collider data. 
In other work{\cite {Rizzo:2005fz}} we have shown that the presence in the action 
of Lovelock invariants can lead to TeV-scale BH in ADD-like models with properties  
that can differ significantly from the usual EH expectations including the 
possibility that BH may be stable in $n$-odd dimensions. The purpose of the present 
paper is to examine the effects of GB terms in the action on TeV scale Schwartzschild-AdS  
BH in the RS model. The possible lifetimes of these BH will be examined and, in 
particular, we focus on the predicted differences between canonical and microcanonical 
thermodynamical descriptions of BH decays and the possibility of long-lived relics. 
 
In the next section we provide a brief overview of the essential aspects of the 
GB-extended RS model necessary for our analysis. In Section 3, we discuss the basic 
properties of the BH in this model and calculate their corresponding production cross 
sections for TeV colliders. We also discuss the differences between the use of the canonical 
ensemble and microcanonical ensemble descriptions for BH Hawking radiation when the 
BH mass is comparable to the 5D effective Planck scale as might be expected at future 
colliders. In Section 4, we present the 
results of a numerical study of BH decay rates in the GB-extended RS model. We perform a 
detailed comparison of the two possible statistical descriptions for BH decay and show 
the sensitivity of these results to variations in the model parameters. The sensitivity 
of BH mass loss to the statistics of the final state particles is also discussed. 
The last section of the paper contains a summary and our conclusions.

\section{RS Background}

The basic ansatz of the RS scenario is the existence of a slice of
warped, Anti-deSitter space bounded by two `branes' which we assume are fixed
at the $S^1/Z_2$ orbifold fixed points, $y=0,\pi r_c$, termed the Planck and TeV
branes, respectively{\cite {RS}}. The 5-d metric describing this
setup is given by the conventional expression
\begin{equation}
ds^2=e^{-2\sigma} \eta_{\mu\nu}dx^\mu dx^\nu-dy^2\,.
\end{equation}
As usual, due to the $S^1/Z_2$ orbifold symmetry one requires $\sigma=
\sigma(|y|)$ and, in keeping with the RS solution, we have $\sigma=k|y|$
with $k$ a dimensionful constant of order the fundamental Planck scale.
As first shown in Ref.{\cite {KKL}} the inclusion of GB terms is the only one that 
does not alter this 
basic setup. Based on the above discussion, the action for the model we consider 
takes the form
\begin{equation}
S=S_{bulk}+S_{branes}\,,
\end{equation}
where
\begin{equation}
S_{bulk}=\int d^5x ~\sqrt {-g} \Bigg[{M^3\over {2}} R-\Lambda_b+
{\alpha M\over {2}} \Big(R^2-4R_{AB}R^{AB}+R_{ABCD}R^{ABCD}\Big)\Bigg]\,,
\end{equation}
describes the bulk with $M$ being the $D=5$ fundamental Planck scale,
$\Lambda_b$ the bulk cosmological constant and $\alpha$ the above mentioned 
dimensionless constant which measures the relative strength of the GB interaction. 
If we anticipate that the GB terms be (mainly) subdominant 
to the EH ones as might arise in some sort of loop perturbative expansion 
then one can argue{\cite {Rizzo:2005fz,Hewett:2005iw}} 
that the natural size of this parameter is $|\alpha| \sim 1/D^2 \sim 1/16\pi^2 \sim$ O(0.01) 
which we will assume 
in what follows. In order to avoid both ghosts and tachyons{\cite {Davoudiasl:2005uu}} 
in both the perturbative graviton KK and radion{\cite {radion} sectors, 
$\alpha$ must be chosen negative. Similarly
\begin{equation}
S_{branes}=\sum_{i=1}^2 \int d^4x~\sqrt {-g_i} \Big ({\cal {L}}_i
-\Lambda_i \Big)\,,
\end{equation}
describes the two branes with $g_i$ being the determinant of the
induced metric and $\Lambda_i$ the associated brane tensions;
the ${\cal {L}}_i$ describe possible SM fields on
the branes. In what follows we will assume as usual for simplicity that the SM fields are
all localized on the TeV brane at $y=\pi r_c$. The (Einstein) equations of motion then 
lead to the following parametric relations:  
\begin{eqnarray}
\Lambda_{Planck}=-\Lambda_{TeV}&=&6kM^3\Big(1- {4\alpha k^2 \over {3M^2}}
\Big)\nonumber \\
6k^2\Big(1-{2\alpha\over {M^2}}k^2\Big) &=& -{\Lambda_b
\over {M^3}}\,,
\end{eqnarray}
where we explicitly see the $\alpha$-dependent modifications to the conventional RS results. 
In the classic RS case, one ordinarily assumes that the ratio $k^2/M^2$ is small 
to {\it avoid} higher 
curvature effects. Here such terms are put in from the beginning and we need no longer 
make this {\it a priori} assumption; we will allow a wide range of $k^2/M^2$ values 
in the analysis that follows. For later convenience we will define the parameter 
$M_*=M\epsilon=Me^{-\pi kr_c}$, which is the warped-down fundamental scale, 
$\sim$ TeV. This parameter in many ways acts in a manner similar to the fundamental $\sim$ TeV 
scale in ADD models when discussing BH production on the RS TeV brane as will be seen in 
the following analysis. 

Since collider experiments with center of mass energies $\sqrt s \sim$ 
TeV are directly probing the mass scale $M$, we might imagine that higher order terms in the 
gravity action may be of some importance when considering RS signatures. Here we examine 
how the GB term will modify RS BH properties.

\section{Black Hole Properties}

How do we describe a Schwartzschild-like BH in $AdS_5$ that is created on the TeV 
brane through SM particle collisions when GB terms are present in the action? In the 
usual 
BH analysis within the ADD framework{\cite {Kanti}}, one employs a modified form of 
the conventional $D$-dimensional Schwartzschild solution{\cite {Wheeler,big}} assuming 
all dimensions are infinitely large. Under the assumption that 
the size of the BH, given by its Schwartzschild radius, $R_s$, is far smaller than the 
size of the compactified dimension, $\pi R_c$, this is a valid approximate description.   
This approximation is seen to hold to a very high degree in the usual ADD models with 
low values of $D$. The analogous requirement in the RS case for a 
TeV brane BH is that $(\epsilon R_s/\pi r_c)^2 <<1$ which is also well satisfied as we 
will see below. A 
further requirement in the RS case is that the bulk is now $AdS_5$ and not Minkowski-like 
as in ADD so that asymptotically flat $D=5$ Schwartzschild-like solutions are not directly 
applicable here. This becomes 
immediately obvious when we think of the curvature of the space-time, $\sim k\epsilon$ 
as measured on the TeV brane, becoming comparable to $R_s^{-1}$. Fortunately, 
the solutions for this asymptotic $AdS$-Schwartzschild BH were found long ago{\cite {big}} 
and such BH have had their properties discussed in some detail in the literature; we 
will make use of these results in the analysis 
that follows. Clearly, RS BH in models with large values of $k^2/M^2$ may differ 
significantly from 
their ADD cousins and this is particularly true when higher order curvature terms are 
present in the action as they are here. 

The first step in our analysis is to determine the relationship between $R_s$, $M_*$ and 
the BH mass $M_{BH}$. Note that it is $M_*$ that appears in these relations and not $M$ 
since all the SM matter is assumed to be on the TeV brane. Following, \eg, Cai{\cite {big}}, 
and employing the definitions above as well 
as the relationship between $k$ and $\Lambda_b$ implied by the Einstein equations of 
motion in Eq.(6), we find that 
\begin{equation}
m=c\Big[x^2+2\alpha +\gamma x^4\Big]\,,
\end{equation}
where we employ the dimensionless quantities $x=M_*R_s$, $m=M_{BH}/M_*$, 
\begin{equation}
\gamma ={k^2\over {M^2}}\Big(1-2\alpha {k^2\over {M^2}}\Big)\,,
\end{equation}
and the numerical factor $c=3\pi^2$. 
Note that $\gamma$ is a measure of the curvature of the $AdS_5$ space; when 
$\gamma \to 0$ the bulk becomes `flat', \ie, Minkowski-like and we recover the 
ADD result for $D=5$ with GB terms in the action. To cover all eventualities we 
will consider the range $10^{-4} \leq \gamma \leq 1$ in our subsequent 
numerical analysis. Since we generally 
want $x(m)$ and not $m(x)$ as above we simply invert the expression in Eq.(7) to give 
\begin{equation}
x^2=(2\gamma)^{-1}\Bigg[-1+\Big[1+4\gamma(m/c-2\alpha)\Big]^{1/2}\Bigg]\,,
\end{equation}
where the sign has been chosen to insure that $x^2\geq 0$. In $D=5$ ADD-like models, to 
insure that $x^2 \geq 0$ one requires that $m \geq 2\alpha c$ (since there $\alpha$ is 
positive) which would indicate the existence of a mass threshold for BH production. Here 
there is no {\it apparent} threshold of this type arising from Eq.(9) since 
$\alpha$ is {\it negative}. Note that when $m\to 0$, $x$ remains {\it finite}; conversely, 
note that $x=0$ corresponds to a {\it negative} value of $m$ since $\alpha <0$. This 
behavior is quite distinct from that in ADD-like models even with Lovelock terms present. 
However, the typical BH radii obtained from this expression are numerically quite comparable 
to those obtained in the usual ADD model so that large BH cross sections should be expected 
at the LHC.

Given the mass-radius relationship we can now ask if the requirement  
$(\epsilon R_s/\pi r_c)^2 <<1$ discussed above is satisfied. Recall that we do not want 
our BH to `see' the fact that it is living in a bounded space which would invalidate our 
solution. Using the notation above 
we can re-write this condition as $(k^2/M^2)(\pi kr_c)^{-2} x^2<<1$; if we take typical 
values of these quantities: 
$k^2/M^2 \lsim$ O(1), $x \sim $O(1) and $kr_c \sim 11$, we see that this constraint is very 
easily satisfied for our range of parameters. 

The BH Hawking temperature can be obtained from the derivative of the metric tensor 
evaluated on the horizon in the usual manner; following, \eg,  Cai{\cite {big}} we obtain
\begin{equation}
T={1\over {2\pi}} {{x+2\gamma x^3}\over {x^2+4\alpha}}\,.
\end{equation}
Since $\alpha$ is negative while $\gamma$ is positive it is clear that $x^2$ must 
be bounded from below {\it if} we demand that the BH Hawking temperature is to remain 
positive, \ie, $x^2 \geq -4\alpha$. Through the 
mass-radius relationship this implies that there 
is a corresponding critical lower bound on the BH mass, $m_{crit}$: 
\begin{equation}
m_{crit}=-6\pi^2 \alpha(1-8\gamma \alpha)\,,
\end{equation}
which is of O(1) when $|\alpha|$ is O(0.01); smaller BH masses lead to negative 
temperatures. Note that the BH temperature is 
infinite precisely at $m=m_{crit}$ instead of at $m=0$ in the usual flat-space EH case. 
This is also unlike the ADD-like models where 
a minimum BH mass also arises from the presence of Lovelock terms in the action 
in odd numbers of extra dimensions. In that case both $x,T\to 0$ at a fixed value of 
$m${\cite {Rizzo:2005fz}} producing the mass threshold. Here, as a function of the BH 
radius, $x$, the temperature starts off infinite at $x^2=-4\alpha$, goes through 
a minimum at some fixed radius and then grows rapidly again as $x \to \infty$ due to the 
finite value of the curvature factor $\gamma$. The threshold temperature behavior in GB 
augmented RS is thus more like that obtained from the traditional $D$-dimensional EH action 
where the temperature diverges as $x \to 0$.

Given the BH mass-radius relationship we can calculate the cross section for BH 
at colliders. The leading approximation for the subprocess cross-section for the 
production of a BH of mass $M_{BH}$ is just its geometric size{\cite {Kanti}}: 
\begin{equation}
\hat \sigma =f\pi R_s^2(\sqrt s=M_{BH})\,,
\end{equation}
where $R_s$ is the 5-dimensional Schwarzschild radius corresponding to
the mass $M_{BH}$ and $f$ is some factor of order unity. The specific range of 
values for this factor has been discussed 
extensively in the literature{\cite {Yoshino:2005hi}}. For our numerical purposes 
we will assume  
$f=1$ in the analysis that follows. Fig.~\ref{fig1} shows the numerical results 
we obtain for the BH cross section as a function of $m/m_{crit}$ for different 
choices of the $\alpha,\gamma$ parameters.  Several features are immediately apparent: 
($i$) BH do not form for masses below $m_{crit}$ as this leads to negative temperatures. 
However, the Schwartzschild radius for a BH of mass $m_{crit}$ is non-zero which 
implies a step-like threshold behavior for the cross section. This is unlike the 
case of the ADD-like models where Lovelock terms produce a very smooth threshold behavior 
for both odd and even numbers of 
dimensions{\cite {Rizzo:2005fz}}. In a more realistic UV completed 
theory it is likely that this threshold is smoothed out to some degree by quantum 
corrections. 
($ii$) The cross section above threshold has an approximately linear 
$m$ dependence; the deviations from linearity are related to the amount of curvature, \ie, 
the size of $\gamma$. For the range of parameters we consider the qualitative 
sensitivity to $\gamma$ is generally rather weak; the numerical scale of the 
cross section is set by $M_*^{-2}$. ($iii$) The overall magnitude of the 
cross section is approximately linear in $-\alpha$. Properties ($i$)-($iii$)  
imply that experimental measurements of the BH cross section at colliders can be 
used to determine the basic GB-extended RS model parameters. 

\begin{figure}[htbp]
\centerline{
\includegraphics[width=8.5cm,angle=90]{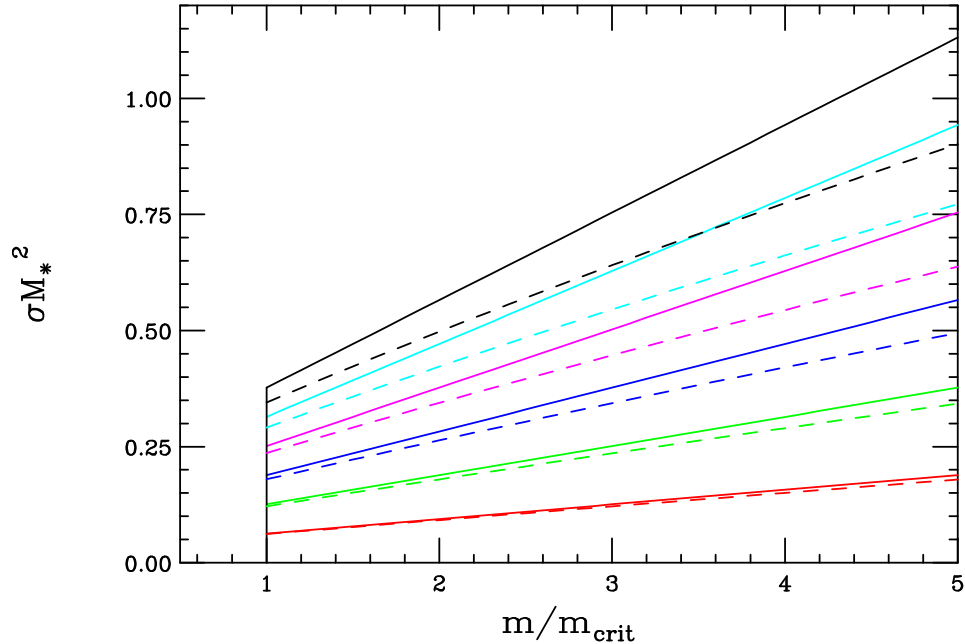}}
\vspace*{0.1cm}
\caption{The scaled BH production cross section as a function of $m/m_{crit}$ for 
$\gamma=10^{-4}$(solid) and 1(dashed) for different values of $\alpha$. From bottom to 
top the curves correspond to $\alpha=-0.005,-0.010...$ increasing in magnitude by  
0.005 for each subsequent curve.}
\label{fig1}
\end{figure}

We now turn to the issue of BH decays through Hawking radiation; for simplicity 
we will ignore the influence 
of possible grey-body{\cite {Kanti,Kanti2}} factors that may be present. It has been shown in 
the flat space case that the inclusion of such factors for brane fields when the GB term 
is present 
does not modify the standard expectation that BH decays to brane fields dominates over bulk 
modes. Cardoso, Cavaglia and Gualtieri{\cite {big}} have, however, shown that bulk 
grey-body factors could upset these expectations when the number of extra dimensions. Assuming 
these results are applicable to the present case we are relatively safe from these effects as 
$n=1$. 

The usual thermodynamical 
description of the Hawking radiation produced by TeV-scale BH decays is via the 
canonical ensemble(CE){\cite {Kanti}} which has been employed in most analyses 
in the literature (in particular, our previous analysis of ADD-like BH).
However, as pointed out by several groups{\cite {Casadio:2001dc}}, 
though certainly applicable to very massive BH, this approach 
does not strictly apply when $M_{BH}/M_*$ is not much greater than O(1) or when the 
emitted particles carry an energy comparable to the BH mass itself due to 
the back-reaction of the emitted particles on the properties of the BH. This certainly 
happens when the resulting overall BH Hawking radiation multiplicity is low.  
In the decay of TeV-scale BH that can be made at a collider, 
the energy of the emitted particles is generally comparable to both $M_*$ as well as 
the mass of the BH itself thus requiring following the MCE treatment. 
In the CE approach the BH is treated as a large 
heat bath whose temperature is not significantly  
influenced by the emission of an individual particle. While this is a very good  
approximation for reasonably heavy BH it becomes worse as the BH mass approaches the 
$M_*$ scale as it does for the case we consider below. 
Furthermore, the BH in an asymptotically flat space (which we can assume is relatively 
good approximation here for small $\gamma$ since the BH Schwarzschild radii, $R_s$, are far 
smaller than the compactification radius as noted above) cannot be in equilibrium with its 
Hawking radiation.  

It has been suggested{\cite {Casadio:2001dc}} 
that all these issues can be dealt with simultaneously 
if we instead employ the correct, \ie, microcanonical ensemble(MCE) approach 
in the statistical mechanics treatment for BH 
decay. As $M_{BH}/M_*$ grows larger, $\gsim 10-20$, the predictions of these two 
treatments will be found to agree, but they differ in the region which is of most  
interest to us since at colliders 
we are close to the BH production threshold where $M_{BH}/M_*$ 
is not far above unity. Within the framework of the 
EH action it has been emphasized{\cite {Casadio:2001dc}} 
that TeV-scale BH lifetimes will be increased by many orders of magnitude when the MCE 
approach is employed in comparison to the conventional CE expectations. This is not due 
to modifications in the thermodynamical quantities, such as the temperature, themselves 
but how they enter the expressions for the rate of mass loss in the decay of the BH. 
Here we will address the issue of how these two statistical 
descriptions may differ in the BH mass range of interest to us when the additional 
higher-curvature GB term is present in the RS action. Here we address these 
CE vs. MCE differences in the case of the RS model with GB terms present. It is important 
to remember that the values of the various thermodynamic quantities themselves , \eg, the 
BH temperature, $T$, are the same in both approaches.

\section{Numerical Results: Mass Loss Rates and Lifetimes}

In order to be definitive in our calculation of the BH Hawking decay rates we follow 
the formalism of 
Hossenfelder{\cite {Casadio:2001dc}}; to simplify our presentation and to focus 
on the GB modifications to RS, as well as potential 
MCE and CE differences, we remind the reader that we will ignore the 
effects due to grey-body factors{\cite {Kanti}} in the present analysis. 
Since bulk decays are expected to be generally sub-dominant and since the only bulk modes 
are gravitons which have KK excitations that are quite heavy in the RS model, $\sim$ TeV and 
the BH masses themselves, we will here 
assume that the only (numerically) important BH decays are into SM brane fields.  
In this approximation the rate for BH mass loss (time here being measured in units of 
$M_*^{-1}$) due to decay into brane fields employing the MCE approach is given by 
\begin{equation}
\Big[{{dm}\over {dt}}\Big]_{brane}=
{{\Omega_{3}^2}\over {(2\pi)^{3}}} ~\zeta(4) x^2
\sum_i ~\int^m_{m_{crit}} ~dy ~(m-y)^{3}N_i\Big [e^{S(m)-S(y)}+s_i\Big]^{-1}\,,
\end{equation}
where, $x=M_*R_s$ as above, $S$ in the corresponding entropy of the BH, 
$i$ labels various particle species which live on the brane and obey 
Fermi-Dirac(FD), Boltzmann(B), or Bose-Einstein(BE) statistics, corresponding to the 
choices of $s_i=1,0,-1$, respectively, with the corresponding 
number of degrees of freedom $N_i$, 
$\zeta$ is the Riemann zeta function with $\zeta(4)=\pi^4/90$, and as usual 
\begin{equation}
\Omega_{d+3}={{2\pi^{(d+3)/2}}\over {\Gamma((d+3)/2))}}\,,
\end{equation}
so that $\Omega_3=4\pi$. To proceed further we need to know the BH entropy $S$; this  
entropy can be calculated using the familiar thermodynamical relation
\begin{equation}
S=\int dx ~T^{-1} {\partial m\over {\partial x}}\,,
\end{equation}
which yields
\begin{equation}
S={{4\pi}\over {3}}c\Big(x^3+12\alpha x+K\Big)\,,
\end{equation}
where $K$ is an integration constant. Since only the entropy difference  
$S(m)-S(y)$ enters into the expression above this constant can be chosen 
arbitrarily within our present analysis. However, perhaps the most convenient choice is to 
choose $K=16(-\alpha)^{3/2}$ such that $S(m_{crit})=0$. In this special 
case $S$ starts at zero when $x^2=-4\alpha$ and monotonically increases as $x$ increases 
since $\partial S/\partial x$ is always positive. The free energy of these BH, $F=m-TS$, is 
also always positive in this case. 

How sensitive is the BH mass loss rate and lifetime to the statistics of the particles in 
the final state? In the 
CE analysis, to be discussed below, this sensitivity is only at the level of $5-10\%$ 
and is due to a simple overall multiplicative factor. To address this issue for 
the MCE approach we display in Fig.~\ref{fig2} the results obtained 
by assuming BH decays to the pure BE, B and FD statistical final states taking $N=60$, 
$\alpha=-0.01$ and $\gamma=10^{-4}$ for purposes of demonstration. While the B and FD 
statistics cases are rather close numerically over the entire mass range, the BE 
choice leads to a more rapid decay and a 
substantially shorter lifetime as can be seen in Fig.~\ref{fig2}. Here we see that 
BH decaying only to fermions may have a lifetime which differs from one decaying only 
into bosons by a few orders of magnitude, $\sim 10^3-10^4$. Note that at larger values of 
$m$, the MCE $\to$ CE limit, all three curves become rather close, at the 
level of $\sim 10\%$ as expected, while still differing at smaller $m$ values. 
\begin{figure}[htbp]
\centerline{
\includegraphics[width=8.5cm,angle=90]{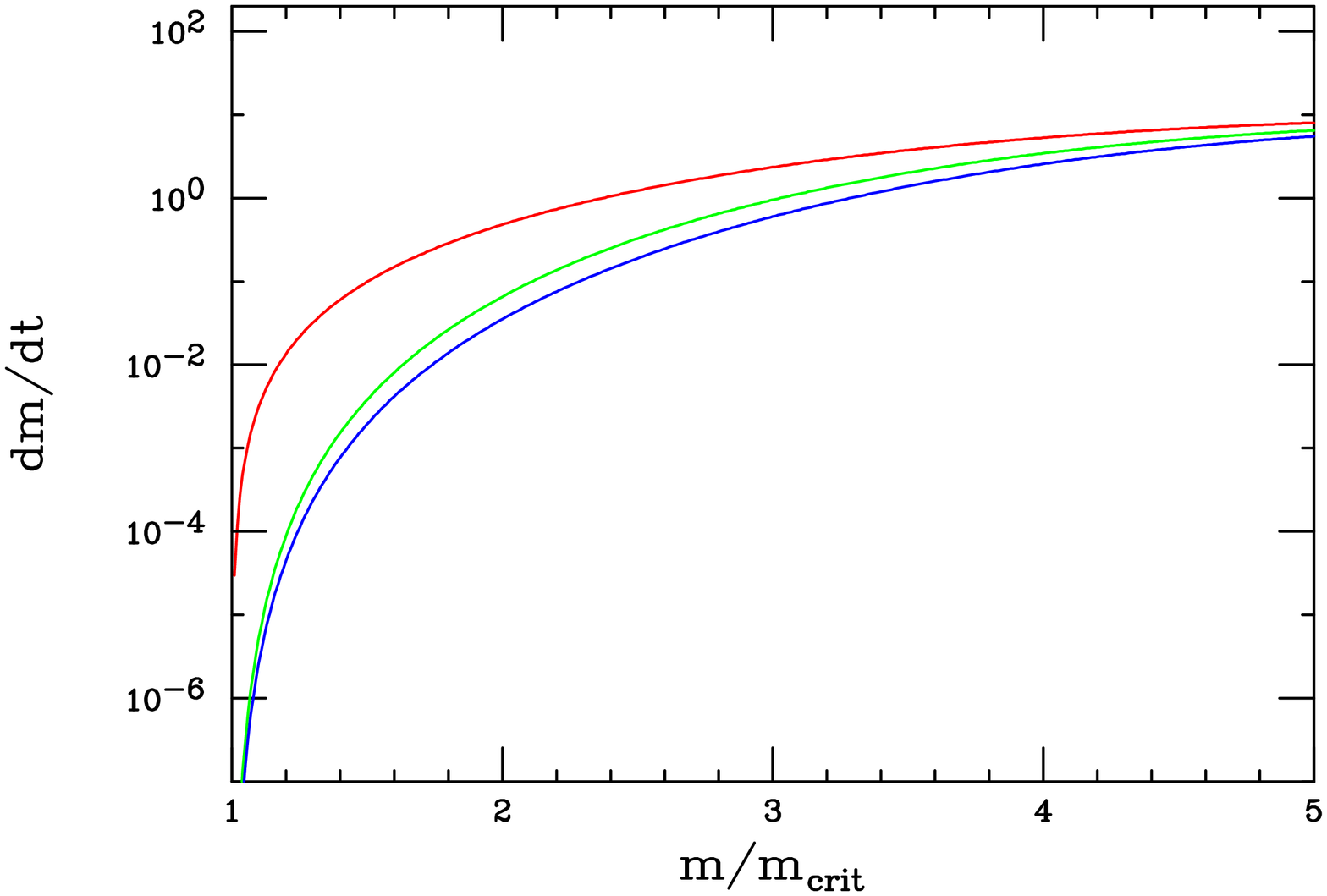}}
\vspace{0.4cm}
\centerline{
\includegraphics[width=8.5cm,angle=90]{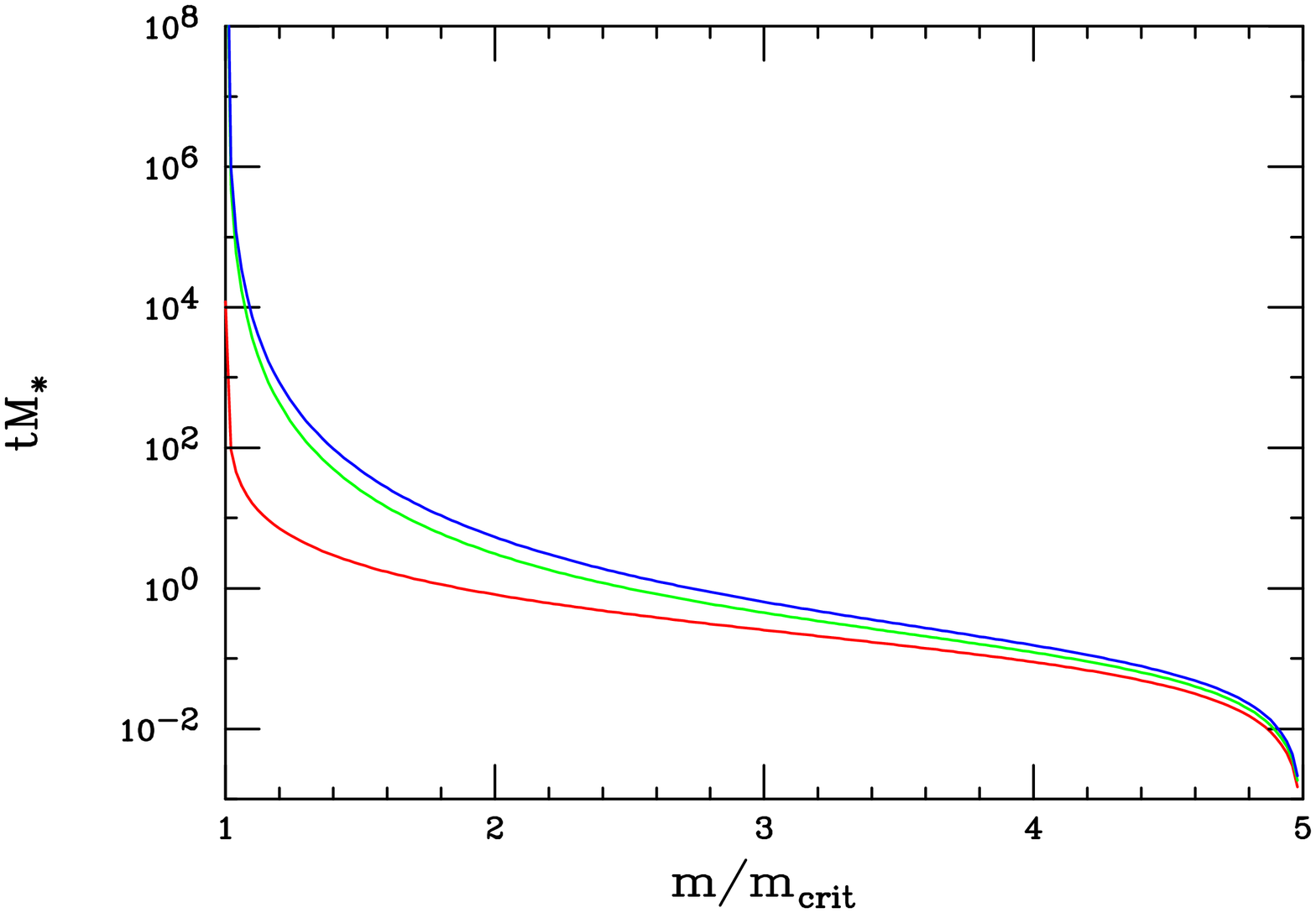}}
\vspace*{0.1cm}
\caption{The BH mass loss rate(top) and lifetime(bottom) in the MCE approach 
as a function of $m/m_{crit}$ assuming an initial value of $m=5m_{crit}$, with $\alpha=-0.01$ 
and $\gamma=10^{-4}$. In the top panel, from top to bottom, 
the curves correspond to the choice of BE, B or FD statistics, respectively. In the 
bottom panel, the order for the statistics choice is reversed.}
\label{fig2}
\end{figure}
Though the detailed
meaning of these these curves will not be discussed until later in this section 
it is clear than these differing statistics can in general be quite important when 
performing the BH mass loss analyses using the MCE approach. 

In the CE case, the expression above simplifies 
significantly as there are no non-trivial integrals remaining. The reason for this 
is that in the CE treatment, the factors $S(m)$ and $S(m-y)$ appearing in the 
exponential factor above are considered nearly the same since backreaction is neglected, 
\ie, one replaces this difference in the MCE expression above by the leading term in 
the Taylor series expansion $S(m)-S(m-y)\simeq y\partial_m S=y/T${\cite {Casadio:2001dc}}. 
Taking the limit $m\to \infty$, \ie, no recoil, and integrating over $y$ then produces 
the familiar CE result:
\begin{equation}
\Big[{{dm}\over {dt}}\Big]_{brane}=
Q~{{\Omega_{3}^2}\over {(2\pi)^{3}}} ~\zeta(4)x^{2}\Gamma(4)T^{4}\,,
\end{equation}
where $Q$ takes the value $\pi^4/90(1,~7\pi^4/720)$ for BE(B, FD) statistics. 
Note that the difference in statistics here in the CE case 
is essentially trivial: just a simple multiplicative factor which is close to unity 
unlike in the MCE approach as seen above where there is a functional difference at 
low $m$ values.
In practical calculations, especially since we are concerned with decays to SM brane 
fields where the numbers of fermionic degrees of freedom 
(48, assuming only light Dirac neutrinos) is somewhat larger 
than the number for bosons (14), the value of $s_i$ 
does not play much of an important role. In our numerical analysis that follows 
we will for simplicity take $s_i=0$ and assume that the number of SM fields is 
60. The reason why this is a good approximation is that ($i$) the SM is mostly 
fermionic and the results for Boltzmann and FD statistics lie rather close to one  
another and ($ii$) the B distribution lies between the BE and FD ones. It would be 
interesting to know how this approximation fares for BH decays in the ADD-like case 
when several Lovelock terms are present in the action simultaneously.

\begin{figure}[htbp]
\centerline{
\includegraphics[width=8.5cm,angle=90]{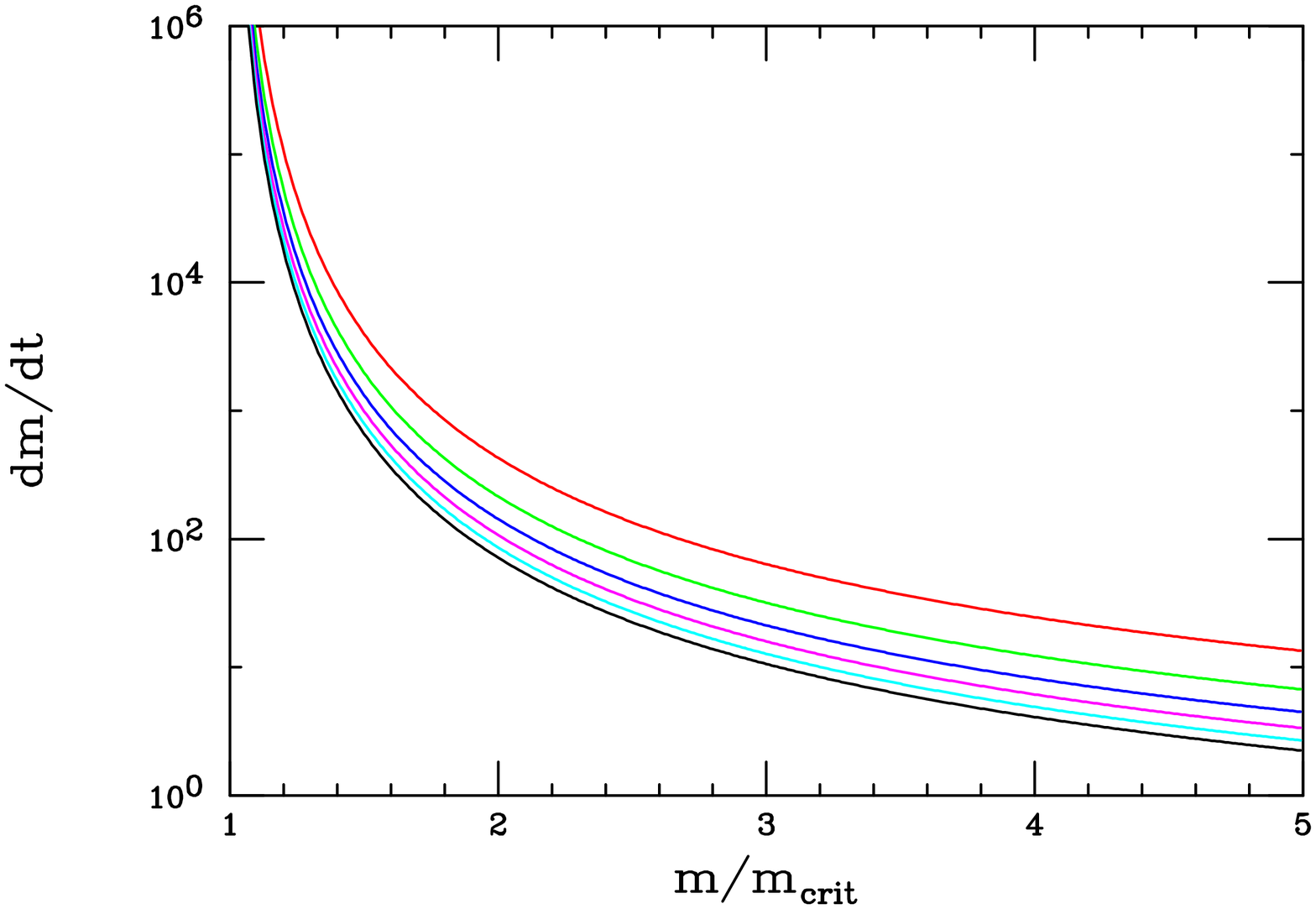}}
\vspace{0.4cm}
\centerline{
\includegraphics[width=8.5cm,angle=90]{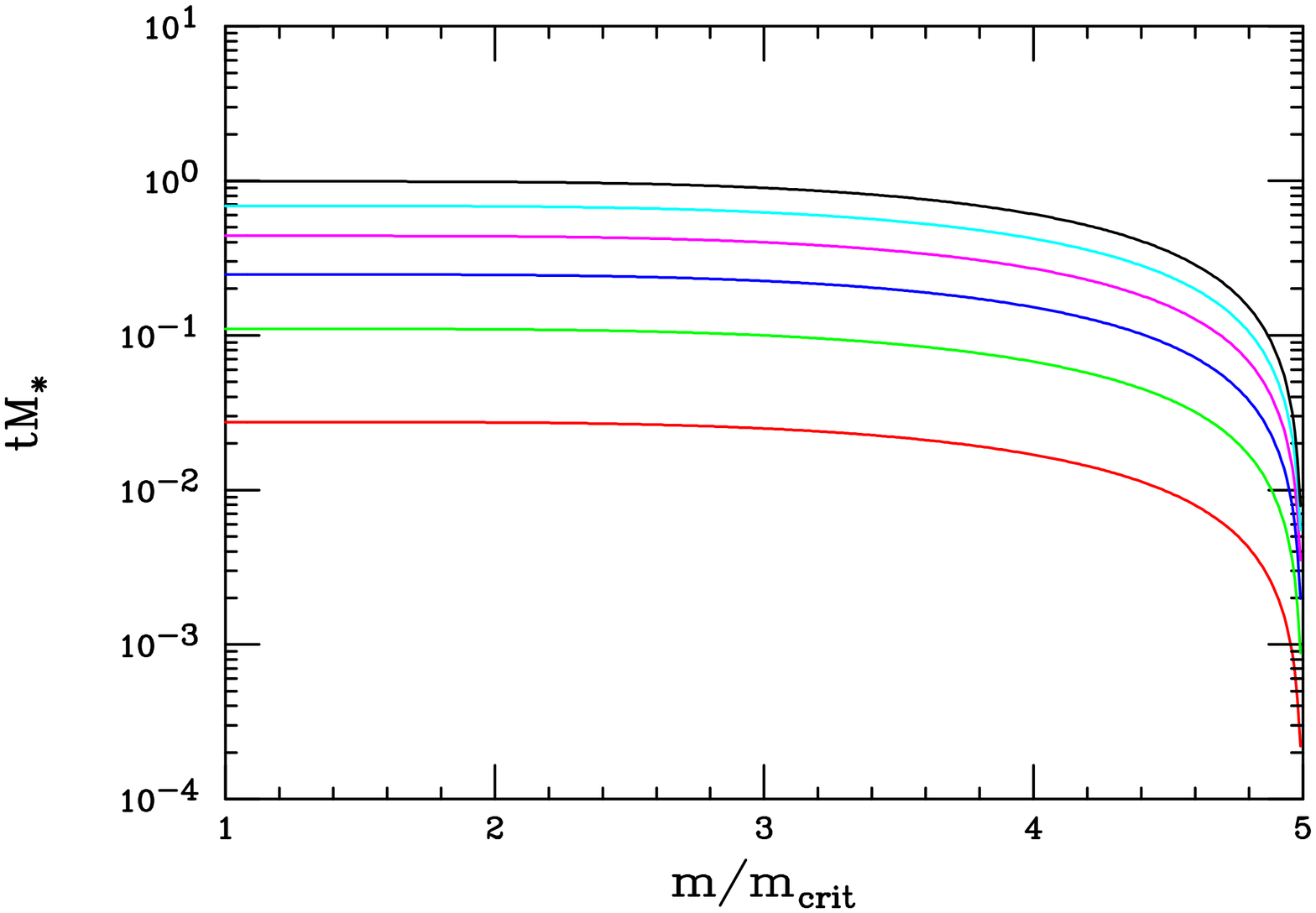}}
\vspace*{0.1cm}
\caption{The BH mass loss rate(top) and lifetime(bottom) as a function of $m/m_{crit}$ 
assuming $\gamma=10^{-4}$ employing the CE scheme and assuming an initial mass $m=5m_{crit}$. 
In the top panel, from top to bottom, 
the curves correspond to the choices $\alpha=-0.005, -0.010,...$. In the 
bottom panel, the order is reversed.}
\label{fig3}
\end{figure}

Given the results above we can now address a number of issues, 
in particular, ($i$) how do BH decay rates and lifetimes depend on the values of 
the parameters $\alpha$ and $\gamma$ and ($ii$) how sensitive are these results to 
the choice of the MCE or CE analysis approach.   
To be specific we first perform our analysis by following the usual CE approach;  
Figs.~\ref{fig3} and ~\ref{fig4} show the results of these specific calculations which 
were performed for two widely different values of $\gamma=10^{-4}$ and $\gamma=1$, 
respectively. Any realistic value for this parameter must lie within this range. 
Assuming a BH with an initial mass of $m=5m_{crit}$, as might be expected at TeV 
colliders, the top panel in both figures shows the rate of mass loss, $dm/dt$, for this 
BH. It is important to note that the mass loss rate is predicted to increase dramatically 
using the CE approach 
as $m\to m_{crit}$ making the BH radiate faster and faster. This is to be expected as 
in the CE approach the mass loss rate is proportional to $T^4$ and $T$ increases dramatically 
as the BH looses mass and gets smaller. This implies that in the CE analysis the final 
state classically stable remnant is reached in a reasonably short amount of time. 
The bottom panel for both $\gamma$ cases shows the time taken for the initial BH 
to radiate down 
to a smaller mass; when $m=m_{crit}$ this is the BH radiative `lifetime', \ie, the time 
taken to decay down to $m_{crit}$. In this CE case this time is short and finite, 
$\sim 0.1 M_*^{-1}$, as might be naively expected; it is not quite clear what 
happens to the remnant without a more complete theory which includes quantum gravity 
contributions but within the present framework we 
are left with a stable remnant. This differs from the case of the ADD-like model with 
Lovelock terms in odd dimensions where the decay down to the finite mass remnant takes 
essentially an infinite amount of time. We further note that the BH specific heat, 
$C=\partial m/\partial T$, here remains negative, as is typical for EH BH, over the relevant 
RS parameter space. This again differs from the ADD-like model with Lovelock terms where 
$C$ can have either sign depending on the values of the Lovelock parameters and the BH 
mass. 

One might wonder if this classical stable BH remnant is unusual and something that happens 
only in extra dimensions (as we already know that it can also occur for flat ADD extra 
dimensions for $n$ odd). Surprisingly, the existence of a classically stable BH remnant is 
a common feature in many models which go beyond the conventional EH 
description including what may happen for a 4-d BH when a renormalization
group running of Newton's constant is employed{\cite {Bonanno}} in order to approximate
leading quantum corrections; such a threshold scenario can also be seen to
occur in theories with a minimum length{\cite {cav}}, in loop quantum
gravity{\cite {loop}} and in non-commutative theories{\cite {nonc}}. Of course, the 
details of this phenomena differ in all these approaches. In, \eg, 
the case of a minimum length, stable remnants occur for all numbers of extra dimensions. 
It is interesting to note that this 
phenomena occurs in all these models where one tries to incorporate 
quantum corrections in some way; though the quantitative nature of such remnants differ 
in detail in each of these models, it would be interesting to learn whether or 
not this is a general qualitative feature of all such approaches. 

\begin{figure}[htbp]
\centerline{
\includegraphics[width=8.5cm,angle=90]{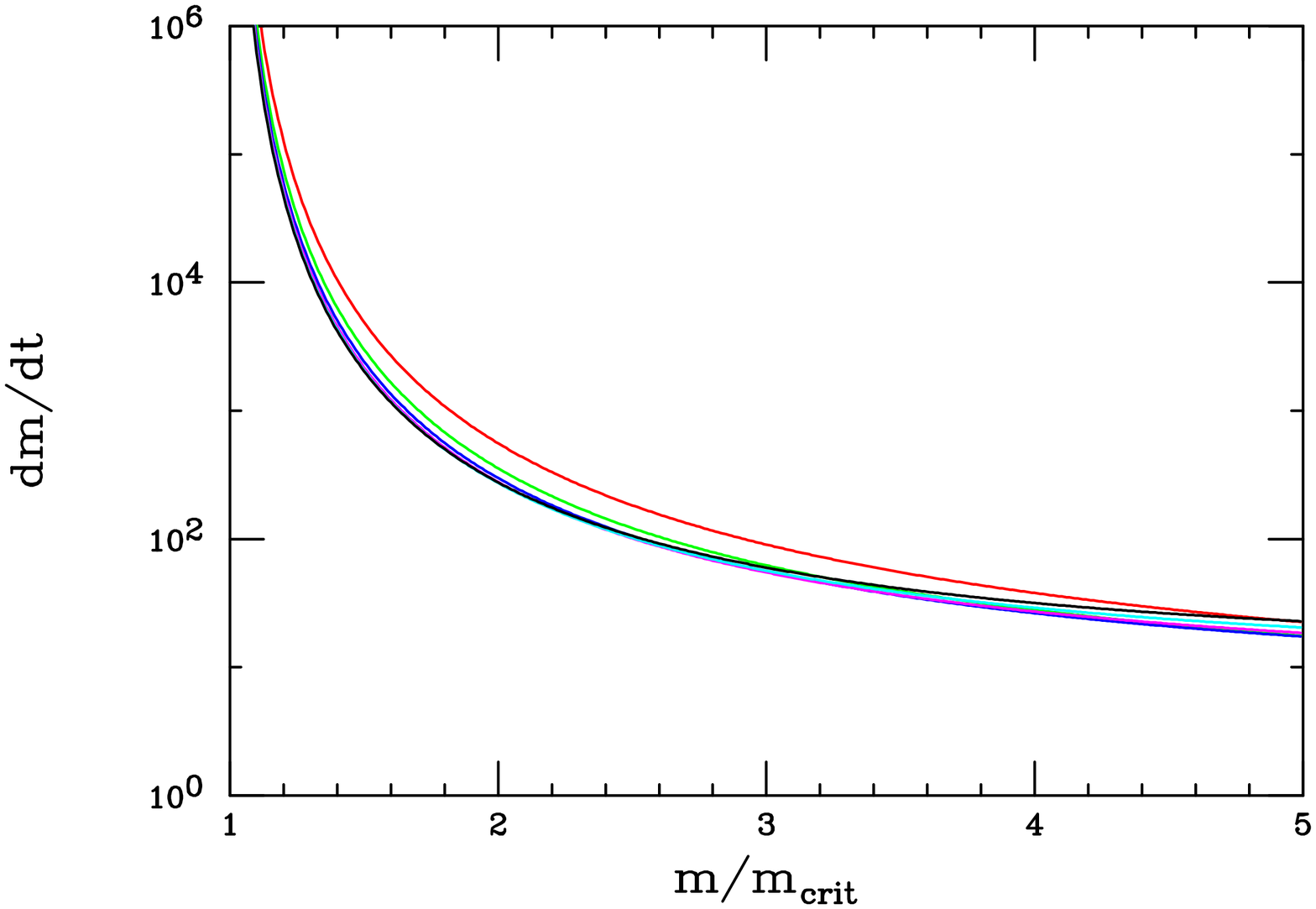}}
\vspace{0.4cm}
\centerline{
\includegraphics[width=8.5cm,angle=90]{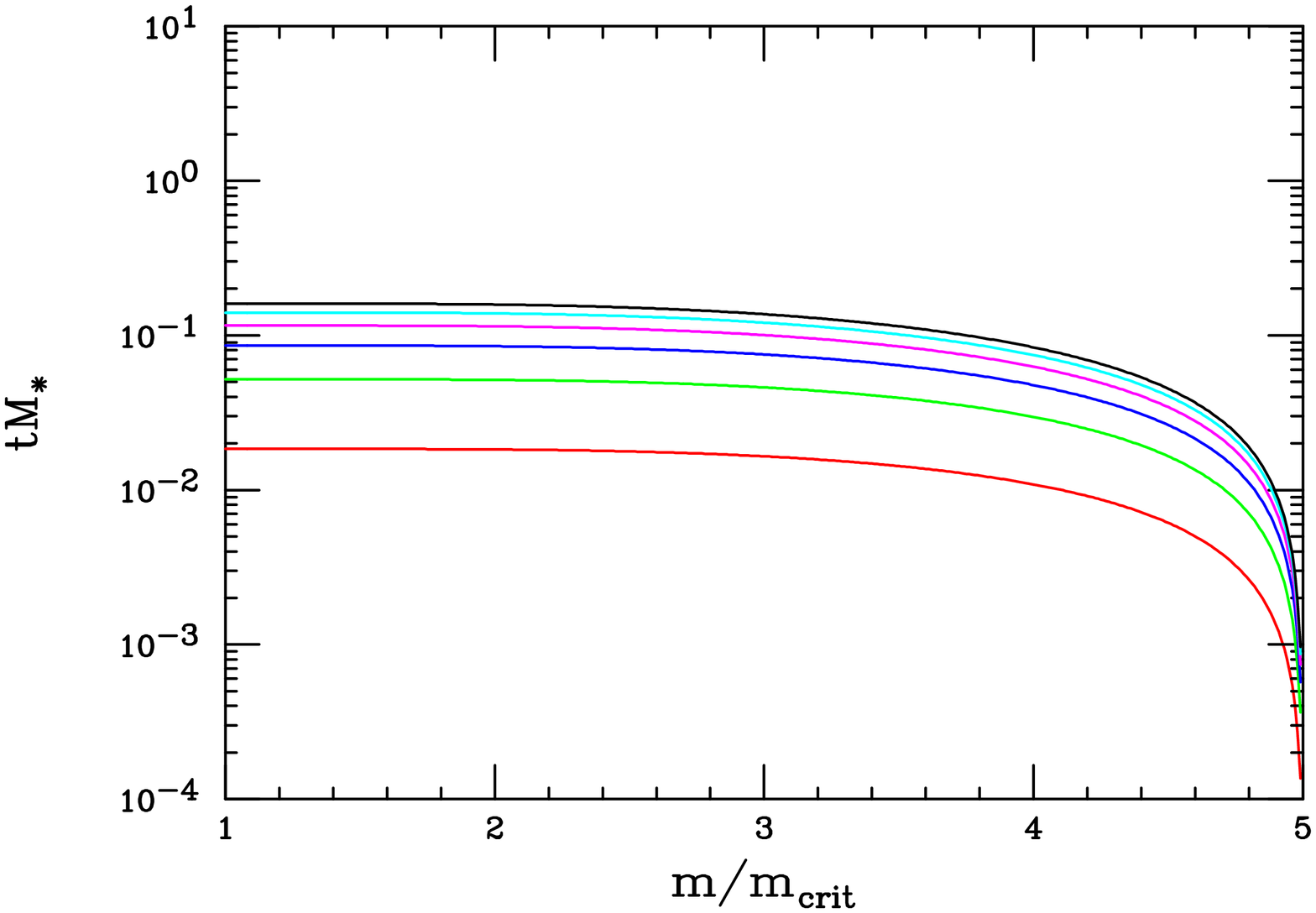}}
\vspace*{0.1cm}
\caption{Same as the previous figure but now assuming $\gamma=1$.}
\label{fig4}
\end{figure}

For fixed $\gamma$, Figs.~\ref{fig3} and ~\ref{fig4} show that both the BH mass loss rate and 
lifetime have only modest dependence on the value of $\alpha$ (for the parameter range 
considered here). The exact $\alpha$ sensitivity is itself, however, dependent on the 
specific  
value of $\gamma$ as a comparison of these two figures show. Generally the BH decay rate 
decreases (and the corresponding lifetime increases) as the magnitude of $\alpha$ is 
raised. 

\begin{figure}[htbp]
\centerline{
\includegraphics[width=8.5cm,angle=90]{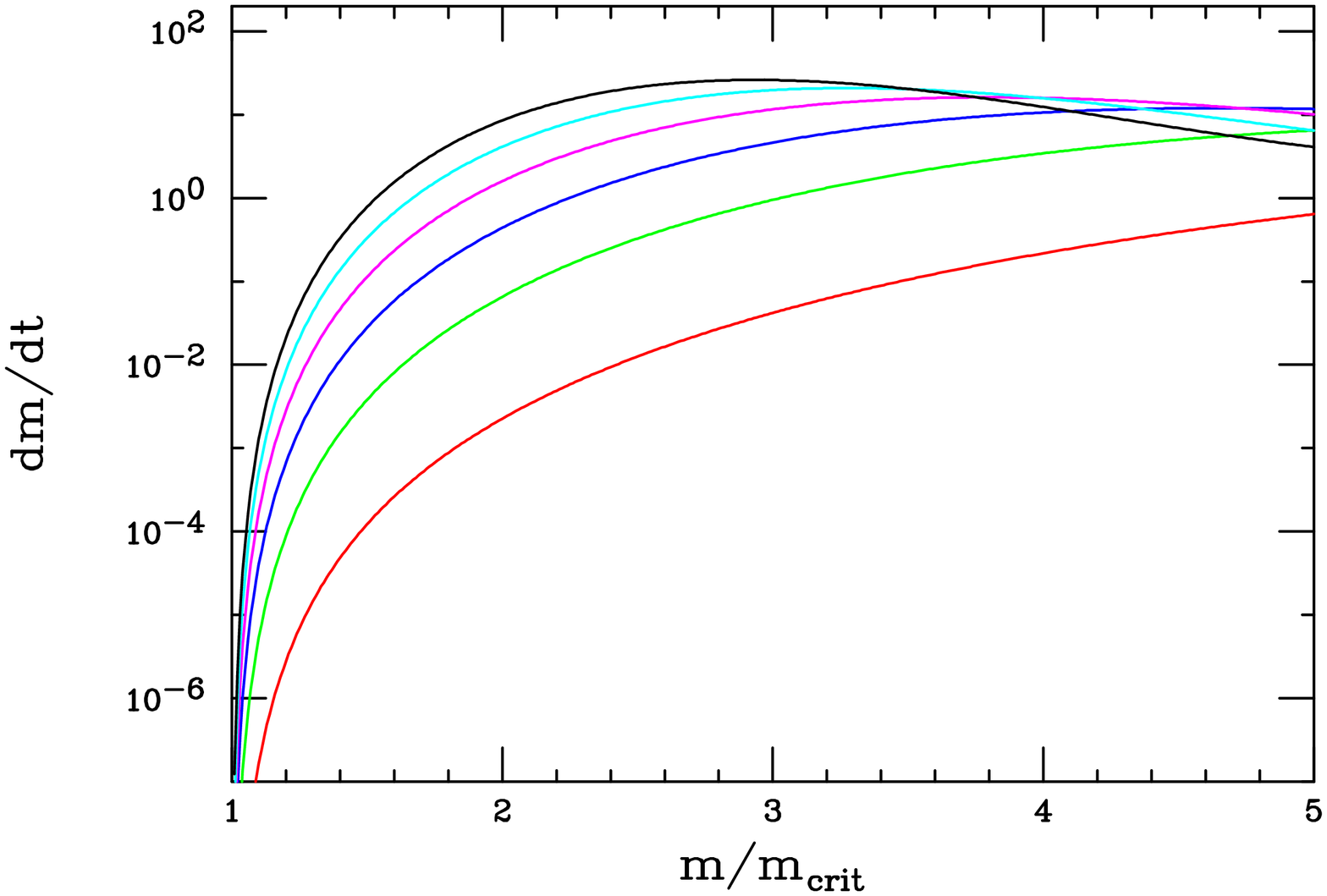}}
\vspace{0.4cm}
\centerline{
\includegraphics[width=8.5cm,angle=90]{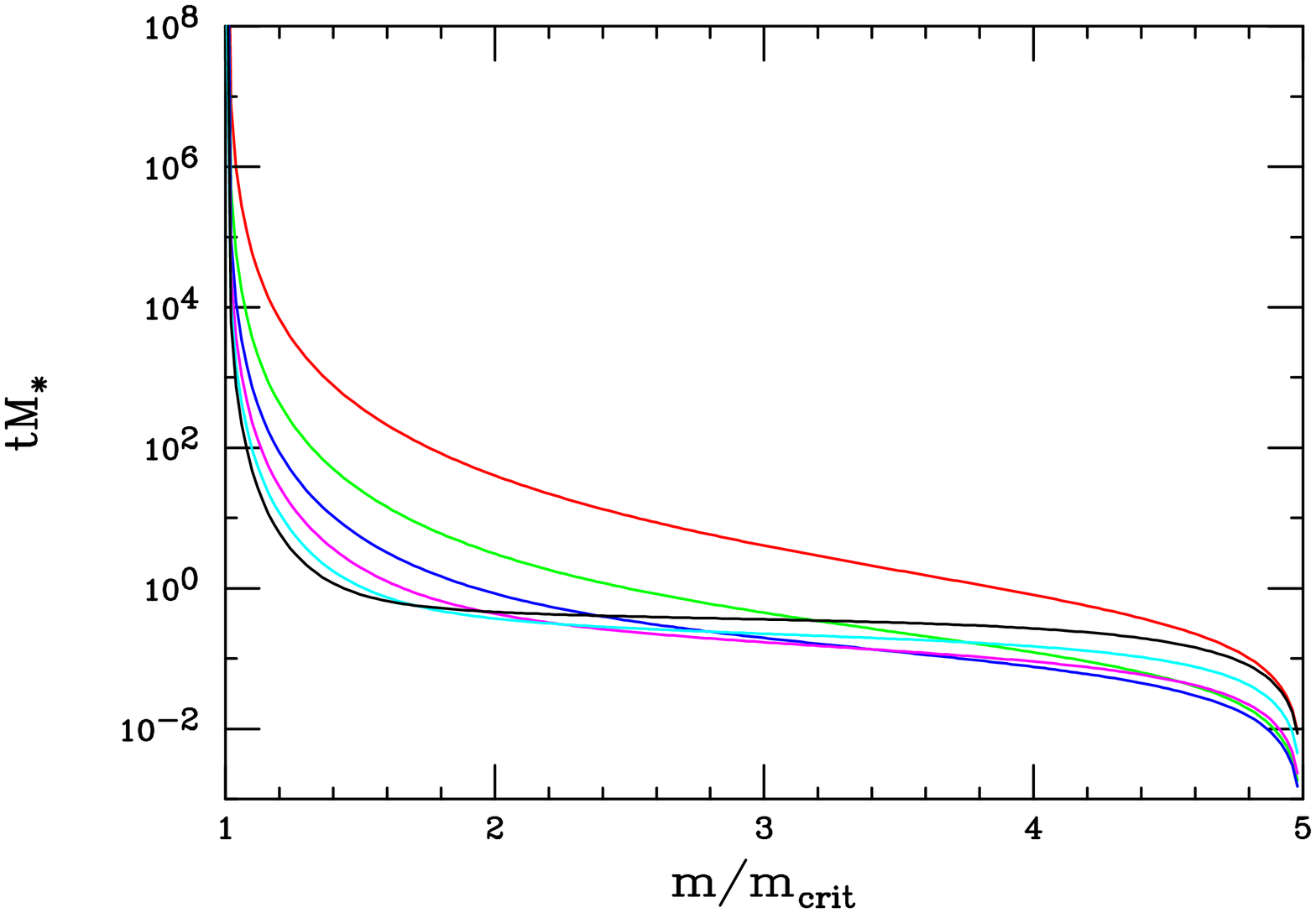}}
\vspace*{0.1cm}
\caption{Same as Fig.~\ref{fig3}, with the opposite ordering of the $\alpha$ dependence 
of the curves, but now using the MCE approach.}
\label{fig5}
\end{figure}

Do these conclusions continue to be valid when we follow the MCE prescription?  
This question can be answered by examining the results presented in 
Figs.~\ref{fig5} and \ref{fig6}. In the MCE case, 
the BH mass loss rate rapidly {\it decreases} as $m$ approaches $m_{crit}$; this is the 
complete opposite behavior from what happens for the CE approach as discussed above. 
Generally we see that there is not a very large $\gamma$ dependence, as we saw before in 
the CE analysis, but now the BH with the  
smaller (in magnitude) value of $\alpha$ leads to the smallest mass loss rates, $dm/dt$. 
This is again the completely opposite behavior to that seen in the CE analysis. We note 
that for $m=5m_{crit}$, the largest value shown, $dm/dt$ is not so different in 
the CE and MCE cases. For much large values of $m \gsim 10-20$ one can check that the 
two sets of calculations yield essentially identical numerical results. 
The fact that $dm/dt$ here decreases as $m \to 0$ implies that following the MCE approach 
a BH lives far longer than if one employs the CE analysis. The bottom 
panels of Figs.~\ref{fig5} and ~\ref{fig6} clearly 
demonstrate this result where we see the vastly longer BH lifetimes obtained for the 
MCE approach. In comparison to CE analysis these BH lifetimes are observed to be 
greater by factors of 
order $\sim 10^{10}$ or so. This result is certainly in line with what might have been 
expected based on the previously performed MCE versus CE analyses performed in the ADD 
model assuming the EH action{\cite {Casadio:2001dc}}. 

\begin{figure}[htbp]
\centerline{
\includegraphics[width=8.5cm,angle=90]{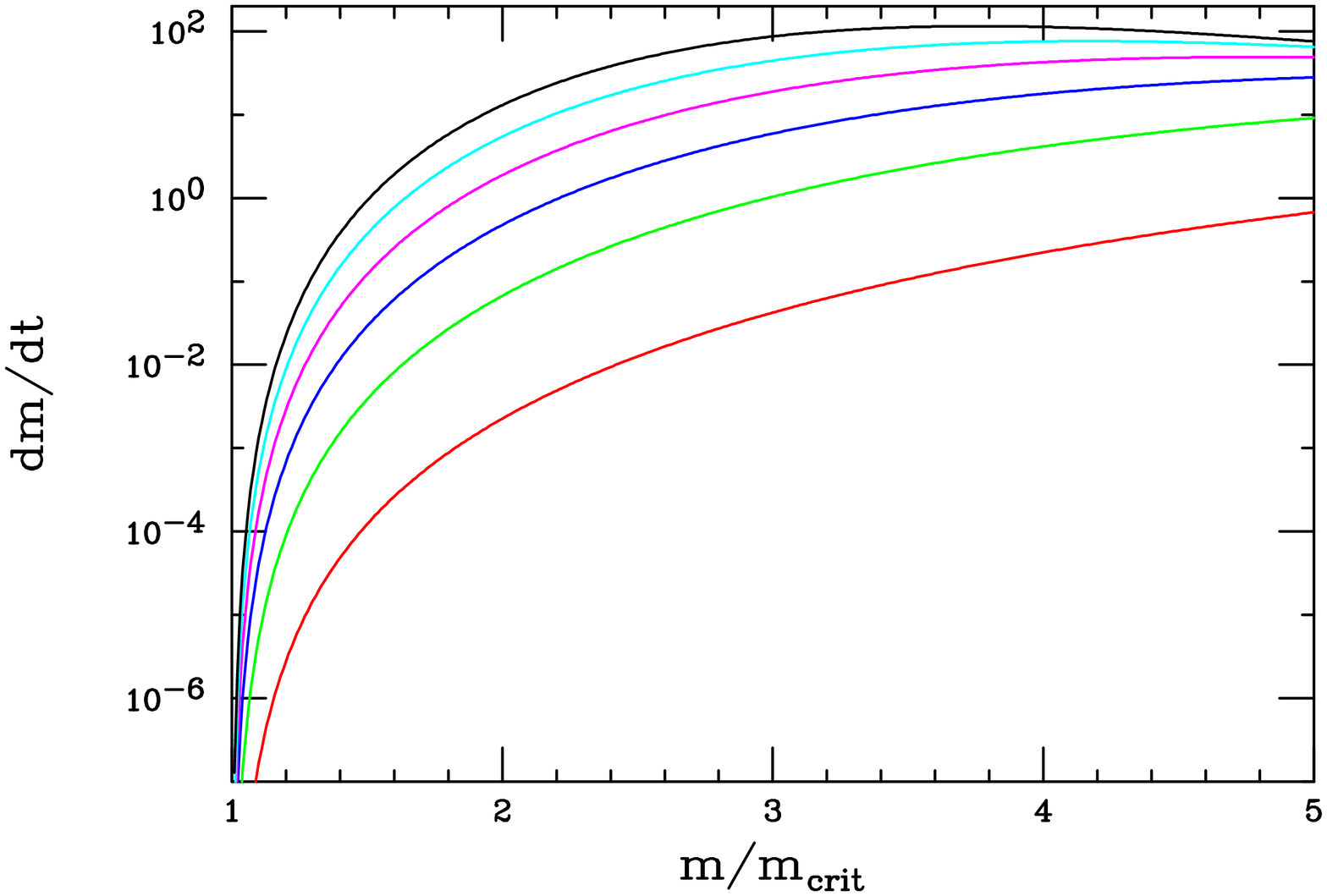}}
\vspace{0.4cm}
\centerline{
\includegraphics[width=8.5cm,angle=90]{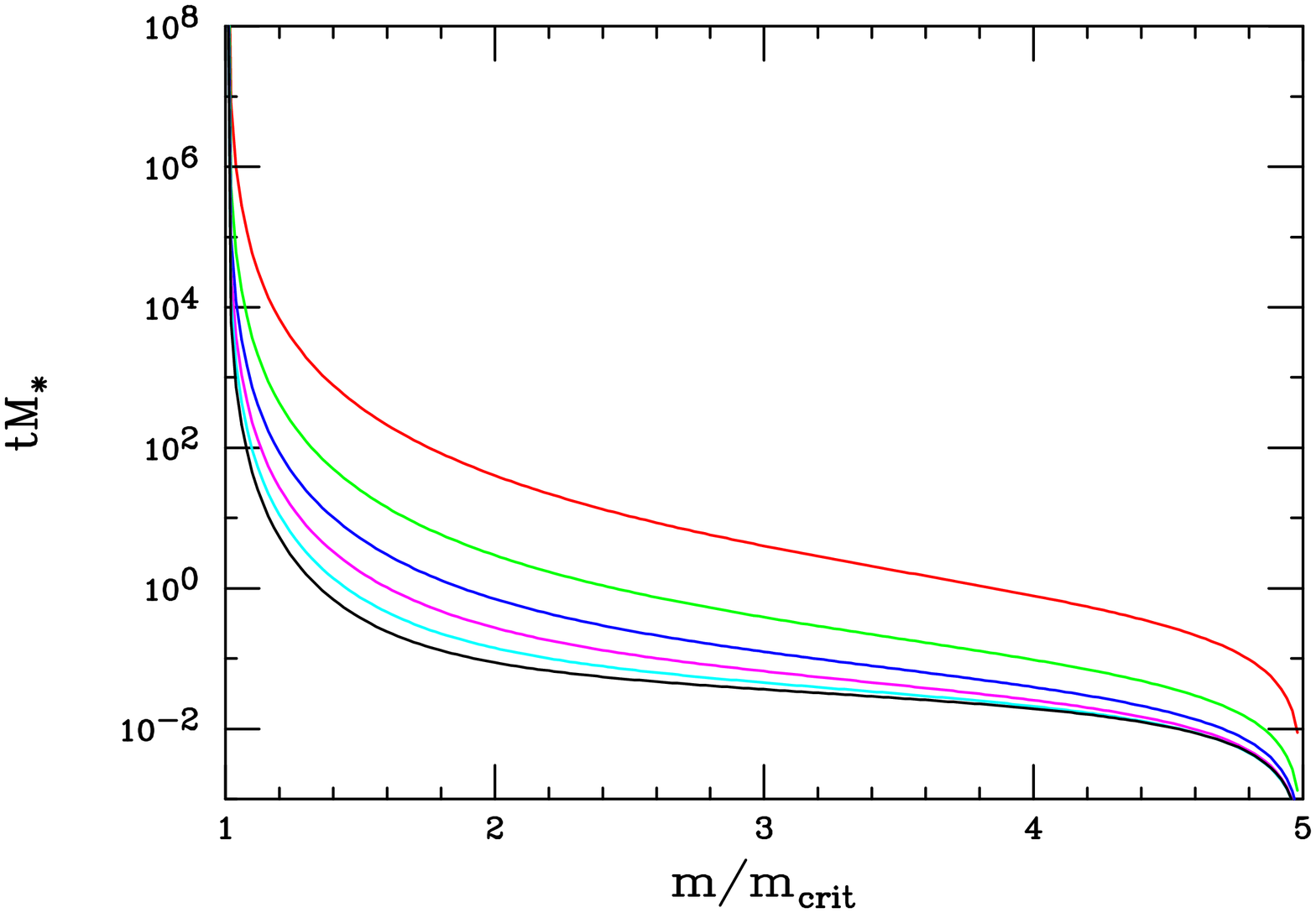}}
\vspace*{0.1cm}
\caption{Same as the previous figure but now for $\gamma=1$.}
\label{fig6}
\end{figure}

The differences in BH mass loss and lifetimes we have obtained here in the RS model due to 
the choice of the MCE vs.CE thermodynamical description demonstrates how much we can learn 
from BH if they are produced at future colliders.

\section{Summary and Conclusions}

In this paper we have analyzed the properties of TeV-scale black holes in the 
Randall-Sundrum model with an extended action containing Gauss-Bonnet terms. 
As discussed above there are many reasons to consider such terms as augmentations to the EH 
action within the RS context: ($i$) model stability, ($ii$) string theory origin and ($iii$) 
good behavior of the resulting equations of motion and graviton propagators. 
In particular we have performed a detailed comparison of BH mass loss rates and 
lifetimes obtained by analyses based on the canonical and microcanonical ensemble 
descriptions. In addition, we have obtained expressions for BH production cross sections 
in these models for future colliders. As in the ADD case, these cross sections are rather 
large yielding millions of events for typical parameter choices and luminosities. 

Overall, the detailed quantitative behavior of BH in the RS model with GB terms in the action 
were found to be quite different than those in ADD-like models with Lovelock terms present. 
Our specific results are as follows: 

($i$) The restriction that the Hawking temperature of a BH be 
positive leads to a lower bound on its radius, $R_s^2 \geq -4\alpha/M_*^2$. 
This implies a corresponding mass threshold, $M_{BH}\geq -6\pi^2 \alpha(1-8\gamma \alpha)
M_*$, which is  
$\sim M_*$ for canonical parameter values, below which BH will not be produced at colliders. 
The resulting production cross section is found to have a step-like behavior, since 
$R_s$ is finite at threshold, and to grow approximately linearly with the BH mass and 
value of $|\alpha|$. 

($ii$) We performed a comparison of the two possible approaches to BH 
thermodynamics based on the canonical and microcanonical ensemble descriptions for the RS 
model. While yielding the same results for large BH masses, as was explicitly verified, 
the two differ in a number of ways when the BH mass is TeV-scale as was considered here.  
First, the BH mass loss rate and lifetime displayed in the CE analysis is well-known to 
display a very trivial dependence on the statistics of the particles into which it decays. 
For light BH we showed that this is not generally true in the MCE approach. For example, 
employing the MCE a BH decaying only to fermions may have a lifetime which differs from 
one decaying only into bosons by several orders of magnitude, $\sim 10^3-10^4$ in the specific 
case examined. For practical calculations involving BH decays only to SM fields the 
statistically weighted mass loss rate was found to be similar to that obtained for decays 
to particles with classical Boltzmann statistics. For light BH the CE and MCE treatments 
were shown to lead to drastically different lifetimes over the entire model parameter 
space. This can be traced to the fact that in the CE analysis the BH mass loss rate, which 
goes as $\sim T^4$, grows rapidly as $m$ decreases since $T$ is then also 
increasing. Given our 
parameter ranges, this then led to BH lifetimes which were typically $\sim 0.1/M_*^2$. On  
the otherhand, the mass loss rate as $m\to m_{crit}$ was found to behave in just the opposite  
manner when the MCE approach was followed. Since the mass loss rate decreases so rapidly 
the corresponding BH lifetimes were found to be enhanced by factors of order $\sim 10^{10}$ 
relative to those of the CE case. This greatly increases the chance of long lived relics 
remaining after the usual BH decay process. 

The differences between ADD-like and RS model BH is rather striking as are those between 
the CE and MCE thermodynamical descriptions. 
It is clear from this discussion that the observation of BH at future TeV colliders will 
provide an important probe of new high scale physics.

\noindent{\Large\bf Acknowledgments}

The author would like to thank J.Hewett and G. Landsberg for discussions related to this work.

%
\def\MPL #1 #2 #3 {Mod. Phys. Lett. {\bf#1},\ #2 (#3)}
\def\NPB #1 #2 #3 {Nucl. Phys. {\bf#1},\ #2 (#3)}
\def\PLB #1 #2 #3 {Phys. Lett. {\bf#1},\ #2 (#3)}
\def\PR #1 #2 #3 {Phys. Rep. {\bf#1},\ #2 (#3)}
\def\PRD #1 #2 #3 {Phys. Rev. {\bf#1},\ #2 (#3)}
\def\PRL #1 #2 #3 {Phys. Rev. Lett. {\bf#1},\ #2 (#3)}
\def\RMP #1 #2 #3 {Rev. Mod. Phys. {\bf#1},\ #2 (#3)}
\def\NIM #1 #2 #3 {Nuc. Inst. Meth. {\bf#1},\ #2 (#3)}
\def\ZPC #1 #2 #3 {Z. Phys. {\bf#1},\ #2 (#3)}
\def\EJPC #1 #2 #3 {E. Phys. J. {\bf#1},\ #2 (#3)}
\def\IJMP #1 #2 #3 {Int. J. Mod. Phys. {\bf#1},\ #2 (#3)}
\def\JHEP #1 #2 #3 {J. High En. Phys. {\bf#1},\ #2 (#3)}


\begin{thebibliography}{99}

\bibitem{RS}
L.~Randall and R.~Sundrum,
Phys.\ Rev.\ Lett.\  {\bf 83}, 3370 (1999)
[arXiv:hep-ph/9905221].

\bibitem{GW}
W.~D.~Goldberger and M.~B.~Wise,
Phys.\ Rev.\ Lett.\  {\bf 83}, 4922 (1999)
[arXiv:hep-ph/9907447].

\bibitem{DHR}
H.~Davoudiasl, J.~L.~Hewett and T.~G.~Rizzo,
Phys.\ Rev.\ Lett.\  {\bf 84}, 2080 (2000)
[arXiv:hep-ph/9909255].

\bibitem{ADD}
N.~Arkani-Hamed, S.~Dimopoulos and G.~R.~Dvali,
Phys.\ Rev.\ D {\bf 59}, 086004 (1999)
[arXiv:hep-ph/9807344] and
Phys.\ Lett.\ B {\bf 429}, 263 (1998)
[arXiv:hep-ph/9803315];
I.~Antoniadis, N.~Arkani-Hamed, S.~Dimopoulos and G.~R.~Dvali,
Phys.\ Lett.\ B {\bf 436}, 257 (1998)
[arXiv:hep-ph/9804398].
                                                                          
\bibitem{Zwiebach}
B.~Zwiebach,
Phys.\ Lett.\ B {\bf 156}, 315 (1985). See also 
D.~G.~Boulware and S.~Deser,
Phys.\ Rev.\ Lett.\  {\bf 55}, 2656 (1985) 
and 
B.~Zumino,
Phys.\ Rept.\  {\bf 137}, 109 (1986).

\bibitem{Mavromatos}
N.~E.~Mavromatos and J.~Rizos,
Phys.\ Rev.\ D {\bf 62}, 124004 (2000)
[arXiv:hep-th/0008074].

\bibitem{Lovelock}
D.Lovelock, J. Math. Phys.{\bf 12}, 498 (1971) .
See also C. Lanczos, Z. Phys. {73}, 147 (1932) ~and Ann. Math. {\bf 39}, 842 
(1938). 

\bibitem{KKL}
J.~E.~Kim, B.~Kyae and H.~M.~Lee,
Nucl.\ Phys.\ B {\bf 582}, 296 (2000)
[Erratum-ibid.\ B {\bf 591}, 587 (2000)]
[arXiv:hep-th/0004005] and 
Phys.\ Rev.\ D {\bf 62}, 045013 (2000)
[arXiv:hep-ph/9912344].

\bibitem{Nojiri}
S.~Nojiri, S.~D.~Odintsov and S.~Ogushi,
Phys.\ Rev.\ D {\bf 65}, 023521 (2002)
[arXiv:hep-th/0108172] and 
JHEP {\bf 0007}, 049 (2000)
[arXiv:hep-th/0006232].

\bibitem{Meissner}
K.~A.~Meissner and M.~Olechowski,
Phys.\ Rev.\ D {\bf 65}, 064017 (2002)
[arXiv:hep-th/0106203] and 
Phys.\ Rev.\ Lett.\  {\bf 86}, 3708 (2001)
[arXiv:hep-th/0009122].

\bibitem{Cho}
Y.~M.~Cho, I.~P.~Neupane and P.~S.~Wesson,
Nucl.\ Phys.\ B {\bf 621}, 388 (2002)
[arXiv:hep-th/0104227].

\bibitem{Cho2}
Y.~M.~Cho and I.~P.~Neupane,
Int.\ J.\ Mod.\ Phys.\ A {\bf 18}, 2703 (2003)
[arXiv:hep-th/0112227].

\bibitem{Neupane}
I.~P.~Neupane,
Class.\ Quant.\ Grav.\  {\bf 19}, 5507 (2002)
[arXiv:hep-th/0106100].


\bibitem{Charmousis}
C.~Charmousis and J.~F.~Dufaux,
arXiv:hep-th/0311267.


\bibitem{Brax}
P.~Brax, N.~Chatillon and D.~A.~Steer,
arXiv:hep-th/0411058.

\bibitem{Gregory:2003px}
  J.~P.~Gregory and A.~Padilla,
  Class.\ Quant.\ Grav.\  {\bf 20}, 4221 (2003)
  [arXiv:hep-th/0304250].


\bibitem{Rizzo:2004rq}
  T.~G.~Rizzo,
  JHEP {\bf 0501}, 028 (2005)
  [arXiv:hep-ph/0412087].

\bibitem{Rizzo:2005fz}
  T.~G.~Rizzo,
  JHEP {\bf 0506}, 079 (2005)
  [arXiv:hep-ph/0503163] and 
  arXiv:hep-ph/0601029.


\bibitem{Hewett:2005iw}
  J.~L.~Hewett, B.~Lillie and T.~G.~Rizzo,
  arXiv:hep-ph/0503178.

\bibitem{Davoudiasl:2005uu}
 For a recent discussion and original references, see,  
  H.~Davoudiasl, B.~Lillie and T.~G.~Rizzo,
  arXiv:hep-ph/0508279.

\bibitem{radion} 
See, for example, 
G.~F.~Giudice, R.~Rattazzi and J.~D.~Wells,
Nucl.\ Phys.\ B {\bf 595}, 250 (2001)
[arXiv:hep-ph/0002178] and  
C.~Csaki, M.~L.~Graesser and G.~D.~Kribs,
Phys.\ Rev.\ D {\bf 63}, 065002 (2001)
[arXiv:hep-th/0008151].


\bibitem{Kanti}
T. Banks and W. Fischler, hep-th/9906038; 
S.~Dimopoulos and G.~Landsberg,
Phys.\ Rev.\ Lett.\  {\bf 87}, 161602 (2001)
[arXiv:hep-ph/0106295];
S.~B.~Giddings and S.~Thomas,
Phys.\ Rev.\ D {\bf 65}, 056010 (2002)
[arXiv:hep-ph/0106219]; 
D.~Stojkovic,
arXiv:hep-ph/0409124;
  S.~B.~Giddings and E.~Katz,
  J.\ Math.\ Phys.\  {\bf 42}, 3082 (2001)
  [arXiv:hep-th/0009176]; 
For a recent review, see 
P.~Kanti,
arXiv:hep-ph/0402168.


\bibitem{Wheeler}
J.~T.~Wheeler,
Nucl.\ Phys.\ B {\bf 273}, 732 (1986) and 
Nucl.\ Phys.\ B {\bf 268}, 737 (1986).
See also, 
B.~Whitt,
Phys.\ Rev.\ D {\bf 38}, 3000 (1988); 
D.~L.~Wiltshire,
Phys.\ Lett.\ B {\bf 169}, 36 (1986) and 
Phys.\ Rev.\ D {\bf 38}, 2445 (1988).
%
See also D.G. Boulware and S.Deser in Ref.{\cite {Zwiebach}}

\bibitem{big}
There is a vast literature on this subject. For a far from exhaustive list
see, for example,
R.~G.~Cai,
Phys.\ Rev.\ D {\bf 65}, 084014 (2002)
[arXiv:hep-th/0109133],
Phys.\ Lett.\ B {\bf 582}, 237 (2004)
[arXiv:hep-th/0311240] and
Phys.\ Rev.\ D {\bf 63}, 124018 (2001)
[arXiv:hep-th/0102113];
  R.~G.~Cai and Q.~Guo,
  Phys.\ Rev.\ D {\bf 69}, 104025 (2004);
  [arXiv:hep-th/0311020].
%
R.~C.~Myers and J.~Z.~Simon,
Phys.\ Rev.\ D {\bf 38}, 2434 (1988);
%
S.~Nojiri, S.~D.~Odintsov and S.~Ogushi,
Phys.\ Rev.\ D {\bf 65}, 023521 (2002)
[arXiv:hep-th/0108172];
%
Y.~M.~Cho and I.~P.~Neupane,
Int.\ J.\ Mod.\ Phys.\ A {\bf 18}, 2703 (2003)
[arXiv:hep-th/0112227] and
Phys.\ Rev.\ D {\bf 66}, 024044 (2002)
[arXiv:hep-th/0202140];
%
T.~Clunan, S.~F.~Ross and D.~J.~Smith,
Class.\ Quant.\ Grav.\  {\bf 21}, 3447 (2004)
[arXiv:gr-qc/0402044];
%
R.~G.~Cai and Q.~Guo,
Phys.\ Rev.\ D {\bf 69}, 104025 (2004)
[arXiv:hep-th/0311020];
%
N.~Okuyama and J.~i.~Koga,
arXiv:hep-th/0501044;
%
J.~Crisostomo, R.~Troncoso and J.~Zanelli,
Phys.\ Rev.\ D {\bf 62}, 084013 (2000)
[arXiv:hep-th/0003271];
%
M.~Cvetic, S.~Nojiri and S.~D.~Odintsov,
Nucl.\ Phys.\ B {\bf 628}, 295 (2002)
[arXiv:hep-th/0112045];
%
R.~Aros, R.~Troncoso and J.~Zanelli,
Phys.\ Rev.\ D {\bf 63}, 084015 (2001)
[arXiv:hep-th/0011097];
%
E.~Abdalla and L.~A.~Correa-Borbonet,
Phys.\ Rev.\ D {\bf 65}, 124011 (2002)
[arXiv:hep-th/0109129];
%
T.~Jacobson, G.~Kang and R.~C.~Myers,
Phys.\ Rev.\ D {\bf 52}, 3518 (1995)
[arXiv:gr-qc/9503020] and
Phys.\ Rev.\ D {\bf 49}, 6587 (1994)
[arXiv:gr-qc/9312023];
%
T.~Jacobson and R.~C.~Myers,
Phys.\ Rev.\ Lett.\  {\bf 70}, 3684 (1993)
[arXiv:hep-th/9305016];
%
S.~Deser and B.~Tekin,
Phys.\ Rev.\ D {\bf 67}, 084009 (2003);
[arXiv:hep-th/0212292].
%
A.~Barrau, J.~Grain and S.~O.~Alexeyev,
Phys.\ Lett.\ B {\bf 584}, 114 (2004)
[arXiv:hep-ph/0311238];
  R.~Konoplya,
  Phys.\ Rev.\ D {\bf 71}, 024038 (2005)
  [arXiv:hep-th/0410057].

                                                                                         


\bibitem{Yoshino:2005hi}
See, for example, 
  H.~Yoshino and V.~S.~Rychkov,
   ``Improved analysis of black hole formation in high-energy particle
   collisions,''
  %
  Phys.\ Rev.\ D {\bf 71}, 104028 (2005)
  [arXiv:hep-th/0503171];
  V.~Cardoso, E.~Berti and M.~Cavaglia,
  Class.\ Quant.\ Grav.\  {\bf 22}, L61 (2005)
  [arXiv:hep-ph/0505125].
  H.~Yoshino, T.~Shiromizu and M.~Shibata,
  arXiv:gr-qc/0508063.

\bibitem{Kanti2}
  J.~Grain, A.~Barrau and P.~Kanti,
  arXiv:hep-th/0509128.

\bibitem{Bonanno}
A.~Bonanno and M.~Reuter,
Phys.\ Rev.\ D {\bf 62}, 043008 (2000)
[arXiv:hep-th/0002196]. See also 
  B.~F.~L.~Ward,
  arXiv:hep-ph/0605054.

 
\bibitem{cav}
See
M.~Cavaglia, S.~Das and R.~Maartens,
Class.\ Quant.\ Grav.\  {\bf 20}, L205 (2003)
[arXiv:hep-ph/0305223] and references therein.
See also,  S.~Hossenfelder,
Phys.\ Lett.\ B {\bf 598}, 92 (2004)
[arXiv:hep-th/0404232].
 
\bibitem{loop}
M.~Bojowald, R.~Goswami, R.~Maartens and P.~Singh, gr-qc/0503041.
 
\bibitem{nonc}
See, for example, 
  P.~Nicolini, A.~Smailagic and E.~Spallucci,
  Phys.\ Lett.\ B {\bf 632}, 547 (2006)
  [arXiv:gr-qc/0510112].
\bibitem{Casadio:2001dc}
See, for example, 
  R.~Casadio and B.~Harms,
  Phys.\ Rev.\ D {\bf 64}, 024016 (2001)
  [arXiv:hep-th/0101154] and 
  Phys.\ Lett.\ B {\bf 487}, 209 (2000)
  [arXiv:hep-th/0004004]; 
  R.~Casadio and B.~Harms,
  Int.\ J.\ Mod.\ Phys.\ A {\bf 17}, 4635 (2002)
  [arXiv:hep-th/0110255]
  P.~Kraus and F.~Wilczek,
  Nucl.\ Phys.\ B {\bf 437}, 231 (1995)
  [arXiv:hep-th/9411219] and 
  Nucl.\ Phys.\ B {\bf 433}, 403 (1995)
  [arXiv:gr-qc/9408003];
  M.~K.~Parikh and F.~Wilczek,
  Phys.\ Rev.\ Lett.\  {\bf 85}, 5042 (2000)
  [arXiv:hep-th/9907001];
  D.~N.~Page,
  Phys.\ Rev.\ D {\bf 13}, 198 (1976);
  E.~Keski-Vakkuri and P.~Kraus,
  Nucl.\ Phys.\ B {\bf 491}, 249 (1997)
  [arXiv:hep-th/9610045];
  S.~Massar and R.~Parentani,
  Nucl.\ Phys.\ B {\bf 575}, 333 (2000)
  [arXiv:gr-qc/9903027];
  T.~Jacobson and R.~Parentani,
  Found.\ Phys.\  {\bf 33}, 323 (2003)
  [arXiv:gr-qc/0302099];
  S.~Hossenfelder, B.~Koch and M.~Bleicher,
  arXiv:hep-ph/0507140;
  B.~Koch, M.~Bleicher and S.~Hossenfelder,
  arXiv:hep-ph/0507138;
  S.~Hossenfelder,
  arXiv:hep-ph/0412265;
  S.~Hossenfelder, S.~Hofmann, M.~Bleicher and H.~Stoecker,
  Phys.\ Rev.\ D {\bf 66}, 101502 (2002)
  [arXiv:hep-ph/0109085];
  V.~Cardoso, M.~Cavaglia and L.~Gualtieri,
  arXiv:hep-th/0512116; 
  V.~Cardoso, M.~Cavaglia and L.~Gualtieri,
  arXiv:hep-th/0512002.


\end{thebibliography}
\end{document}